%Paper: hep-th/9508144
%From: David A. Lowe <lowe@tpau.physics.ucsb.edu>
%Date: Fri, 25 Aug 95 17:11:43 -0700
%Date (revised): Tue, 5 Sep 95 10:28:04 -0700

\input harvmac
\noblackbox
% Poor man's Blackboard Bold characters often used :
\def\inbar{\,\vrule height1.5ex width.4pt depth0pt}
\def\IB{\relax{\rm I\kern-.18em B}}
\def\IC{\relax\hbox{$\inbar\kern-.3em{\rm C}$}}
\def\ID{\relax{\rm I\kern-.18em D}}
\def\IE{\relax{\rm I\kern-.18em E}}
\def\IF{\relax{\rm I\kern-.18em F}}
\def\IG{\relax\hbox{$\inbar\kern-.3em{\rm G}$}}
\def\IH{\relax{\rm I\kern-.18em H}}
\def\II{\relax{\rm I\kern-.18em I}}
\def\IK{\relax{\rm I\kern-.18em K}}
\def\IL{\relax{\rm I\kern-.18em L}}
\def\IM{\relax{\rm I\kern-.18em M}}
\def\IN{\relax{\rm I\kern-.18em N}}
\def\IO{\relax\hbox{$\inbar\kern-.3em{\rm O}$}}
\def\IP{\relax{\rm I\kern-.18em P}}
\def\IQ{\relax\hbox{$\inbar\kern-.3em{\rm Q}$}}
\def\IR{\relax{\rm I\kern-.18em R}}
\font\cmss=cmss10 \font\cmsss=cmss10 at 7pt
\def\IZ{\relax\ifmmode\mathchoice
{\hbox{\cmss Z\kern-.4em Z}}{\hbox{\cmss Z\kern-.4em Z}}
{\lower.9pt\hbox{\cmsss Z\kern-.4em Z}}
{\lower1.2pt\hbox{\cmsss Z\kern-.4em Z}}\else{\cmss Z\kern-.4em Z}\fi}

%
% Journals
%

\def\NP{{\it Nucl. Phys.\ }}

\def\PL{{\it Phys. Lett.\ }}

%
%Fonts
%
\font\tiau=cmcsc10
%
%References
%
\lref\seiberg{N. Seiberg, ``Observations on the Moduli Space of
Superconformal Field Theories", {\it Nucl. Phys. }{\bf B303} (1988) 286.}
\lref\aspmor{P. Aspinwall and D. Morrison, ``String Theory on K3
surfaces", to appear in ``Essays on Mirror Manifolds 2'', hep-th/9404151.}
\lref\walton{A nice pedagogical discussion of string theory on
K3 surfaces can be found in M. Walton, ``Heterotic
String on the simplest Calabi-Yau manifold and its orbifold limits'',
{\it Phys. Rev.} {\bf D37} (1988) 377.}
\lref\aspinpt{P. Aspinwall, ``Enhanced Gauge Symmetries and K3
Surfaces", hep-th/9507012.}
\lref\chpol{S. Chaudhuri and J. Polchinski, ``Moduli Space of CHL Strings'',
hep-th/9506048.}
\lref\chl{S. Chaudhuri, G. Hockney and J. Lykken, ``Maximally Supersymmetric
String Theories in $D<10$'', to appear in {\it Phys. Rev. Lett.},
hep-th/9505054. S. Chaudhuri, Talk given at Strings'95 conference, March 1995,
USC.}
\lref\nikulin{V.V. Nikulin, ``Finite Automorphism Groups of Kahler K3
Surfaces'', {\it Trans. Moscow Math. Soc.} {\bf 2} (1980) 71.}
\lref\hullt{C. Hull and P. Townsend, ``Unity of Superstring Dualities,''
hep-th/9410167, \NP {\bf B438} (1995) 109.}
\lref\witt{E. Witten,``String Theory Dynamics in Various Dimensions,''
hep-th/9503124, \NP {\bf B443} (1995) 85.}
\lref\hs{J. A. Harvey and A. Strominger, ``The Heterotic String is a
Soliton,'' to appear in \NP B, hep-th/9504047.}
\lref\sens{A. Sen, ``String-String Duality Conjecture in Six Dimensions and
Charged Solitonic Strings'', hep-th/9504027.}
\lref\fhsv{S. Ferrarra, J. Harvey, A. Strominger and C. Vafa, ``Second
Quantized Mirror Symmetry'', hep-th/9505162.}
\lref\schsenb{J. Schwarz and A. Sen, ``The Type IIA Dual of the
Six-Dimensional CHL Compactification,'' hep-th/9507027.}
\lref\aspin{P. Aspinwall, Lecture given at Conference on
S-duality and Mirror Symmetry, Trieste, June 5-9, 1995;
``Some Relationships Between Dualities in String Theory'', Cornell
preprint CLNS-95/1359, to appear.}
\lref\witstr{E. Witten, ``Some Comments on String Dynamics'', hep-th/9507121.}
\lref\vafwit{C. Vafa and E. Witten, ``Dual String Pairs with N=1
and N=2 Supersymmetry in Four Dimensions'', hep-th/9507050.}
\lref\senvaf{A. Sen and C. Vafa, ``Dual Pairs of Type II String
Compactification'', hep-th/9508064.}
\lref\neim{H. Niemeier, {\it Jour. Number Theory} {\bf 5} (1973) 142.
J. Conway and M. Sloane, {\it Jour. Number Theory} {\bf 15} (1982) 83.}
\lref\lsw{W. Lerche, A. N. Schellekens, and N. P. Warner, ``Lattices
and Strings'', {\it Phys. Rep.} {\bf 177} (1989) 1.}
\lref\narain{K.S. Narain, \PL {\bf 169B} (1986) 41; K.S. Narain,
M.H. Sarmadi and E. Witten, \NP {\bf B279} (1987) 369.}
\lref\hls{J.A. Harvey, D.A. Lowe and A. Strominger, ``N=1 String Duality'',
hep-th/9507168.}
\lref\vafawit{C Vafa and E. Witten, ``Dual String Pairs with N=1 and
N=2 Supersymmetry in four dimensions'', hep-th/9507050.}
\lref\duff{M. Duff, ``Strong/Weak Coupling Duality from the Dual String'',
hep-th/9501030, \NP {\bf B442} (1995) 47.}
\lref\senvafa{A. Sen and C. Vafa, ``Dual Pairs of Type II String
Compactification'', hep-th/9508064.}

%
%Title page
%
\baselineskip 12pt
\Title{\vbox{\baselineskip12pt
\hbox{NSF-ITP-95-76}\hbox{UCSBTH-95-23} \hbox{\tt hep-th/9508144} }}
{\vbox{\hbox{\centerline{\bf TYPE IIA-HETEROTIC DUALS WITH MAXIMAL
SUPERSYMMETRY }}}}
\centerline{\tiau Shyamoli Chaudhuri\footnote{$^\dagger$}{sc@itp.ucsb.edu}}
\vskip .1in
\centerline{\it Institute for Theoretical Physics}
\centerline{\it University of California}
\centerline{\it Santa Barbara, CA 93106-4030}
\vskip .1in
\centerline{\tiau and}
\vskip .1in
\centerline{\tiau David A. Lowe\footnote{$^*$}{lowe@tpau.physics.ucsb.edu}}
\vskip.1in
\centerline{\it Department of Physics}
\centerline{\it University of California}
\centerline{\it Santa Barbara, CA 93106-9530}
\vskip .5cm
\noindent
Using finite abelian automorphism groups of $K3$ we construct orbifold
candidates for Type IIA-heterotic dual pairs with maximal supersymmetry
in six and lower dimensions. On the heterotic side, these results
extend the series of known reduced rank theories with maximal
supersymmetry. The corresponding Type IIA theories generalize the
Schwarz and Sen proposal for the dual of the simplest
of the reduced rank theories constructed as a novel Type IIA $\IZ_2$ orbifold.

\Date{August, 1995}
%\draftmode

\newsec{Introduction}

There is growing evidence for an exact strong-weak coupling
duality relating the Type IIA string compactified on a K3 surface and
the heterotic string compactified on $T^4$
\refs{\seiberg \aspmor \hullt \witt \sens \hs{--}\aspinpt}.
The resulting N=2 theory in six
dimensions has $24$ massless vector multiplets at generic points
in the moduli space.

Recently, it has been observed that there are additional heterotic
6d $N=2$ and 4d $N=4$ theories on the heterotic side with fewer massless
vector multiplets at generic points \refs{\chl,\chpol}. It is natural
therefore to ask whether string duality extends to other maximally
supersymmetric theories.

This is the question we will address in this paper. Using
orbifold techniques \fhsv, we will construct dual Type II-heterotic
theories applying the duality dictionary provided by the
soliton string construction of \refs{\sens,\hs}.
Since we are interested in compactifications with maximal supersymmetry
on the heterotic side, we must consider automorphisms of $K3$ which
preserve the holomorphic two-form. These are known as symplectic
automorphisms. The finite abelian symplectic automorphism groups
of Kahler $K3$ surfaces have been classified by Nikulin \nikulin.
Aspinwall \aspin\ has already pointed out the close correspondence between
the dimensions of the moduli spaces of the distinguished $K3$ surfaces
with automorphism group $G$, studied in \nikulin, and the
rank reduction of the 4d maximally supersymmetric heterotic string
models constructed in \chl. In this paper, we will
construct a Type II-heterotic dual pair for each abelian group
in Nikulin's classification. For the case of the cyclic
groups $\IZ_n$ ($n=2,\cdots,8$) we obtain dual pairs in six
dimensions with $N=2$ supersymmetry. In the other cases $G$
is a product of cyclic groups, and we obtain dual pairs
in three, four and five dimensions.

\newsec{Nikulin's Classification and the Dual Pair Construction}

In the following, we will use the conjectured strong-weak coupling
duality between the
heterotic string compactified on a three-torus, and $d=11$ supergravity
compactified on $K3$ \refs{\hullt, \witt}, to construct dual theories
in lower dimensions. The strategy will be to compactify further on
a circle (or more generally a torus), and then apply the orbifold
construction. The symmetries we will utilize combine one of Nikulin's
abelian automorphisms with a translation on the circle. Provided
these symmetries have no fixed points on the supergravity side, we
may apply the duality dictionary of \refs{\sens,\hs} to map the action
of the symmetry onto the heterotic side.

It is important to bear in mind that the compactification of
$d=11$ supergravity may alternatively be described
in an appropriate region of the moduli space as a
compactification of the Type IIA string,
which involves non-trivial RR background
fields \refs{\fhsv, \schsenb}. The shift in the $S^1$ of
$d=11$ supergravity corresponds to giving a Wilson line expectation
value to the RR $U(1)$ field of the ten-dimensional Type IIA string.
In the examples discussed below, the perturbative spectrum
in the untwisted sector is insensitive to the presence of the
Wilson line, however it does affect the spectrum in the twisted sector.
Further discussion of this point may be found in \schsenb.

As Nikulin has shown \nikulin, the action of the automorphism group on the
$K3$ generates a corresponding action on the cohomology.
These automorphisms leave the self-dual forms invariant (we choose an
anti-self-dual Riemann curvature such that the $(0,2)$, the $(2,0)$ and
the Kahler forms are self-dual \walton).
For a $K3$ surface, $H^2(K3, \IZ)$  with its intersection form is isomorphic
to an even self-dual lattice with signature $(19,3)$.
The action of the
group on the cohomology of $K3$ is therefore isomorphic to an automorphism of
this even self-dual lattice.
According to \refs{\sens,\hs}, this lattice is to be identified
with the Narain lattice \narain\
of the heterotic string compactified on a three-torus.
The three right-moving heterotic string coordinates
correspond to the three self-dual elements of $H^2(K3)$, while
nineteen of the left-moving heterotic string coordinates correspond to the
nineteen anti-self-dual elements of $H^2(K3)$.
Invariance of the right-moving coordinates (up to shifts) under this
automorphism means that supersymmetry is preserved. As mentioned above,
these automorphisms of $K3$ are to be combined with shifts on the
additional torus $T^m$ in order that the symmetry be freely acting on
the supergravity side. On the heterotic side, we will start with a lattice
\eqn\hetlat{
\Gamma^{(19+m,3+m)} = \Gamma^{(19,3)} \oplus \Gamma^{(m,m)}~,
}
where the $\Gamma^{(m,m)}$ factor corresponds to the torus.
The shifts on the supergravity side may be identified with
corresponding shifts on $\Gamma^{(m,m)}$.

At this point we recall some useful facts about self-dual lattices.
The euclidean even self-dual lattices up to dimension 24 (Niemeier lattices)
have been classified \neim . They include 23 Lie algebra lattices and
the famous Leech lattice which is unique in having no vectors of norm 2.
These lattices and some useful tricks for lattice manipulation are
described in appendix A of \lsw . The Niemeier classification is a
nice starting point for constructing symplectic asymmetric orbifolds.

In the following subsections we consider in turn each of the abelian
groups appearing in Nikulin's
classification, and construct dual pairs of maximally supersymmetric
heterotic and Type II theories. In each subsection we will describe the
action of the symmetry on $K3$ and the cohomology, and then discuss
the action on a corresponding Narain lattice
of the heterotic string compactification, in some cases explicitly
constructing such lattices at enhanced symmetry points.
First we consider the cyclic groups $\IZ_n$ $(n=2,\cdots,8)$
that appear in Nikulin's list.
The $d=11$ supergravity theory is compactified on
\eqn\xfive{
X= {{K3\times S^1} \over {\IZ_n}}~,
}
where the symmetry acts on the circle as a $\IZ_n$ shift.
This yields a Type II supergravity theory in six dimensions
with $N=2$ supersymmetry.
Later we will consider the noncyclic cases. In these cases we will
consider orbifold compactifications down to lower dimensions,
with shifts in the
additional torus to ensure the symmetry is freely acting on $K3 \times T^m$,
where $m=2,3,4$.

The classical moduli spaces of $K3$ surfaces admitting abelian
automorphisms are studied in \nikulin. For each of the orbifold
theories we study, a nontrivial moduli space will therefore exist.
Since the theories are maximally supersymmetric, the local structure
of the moduli space is determined by the number of vector
multiplets and takes the form
\eqn\nnspace{
{\cal M} = {{SO(20+m-r,4+m; \IR) }\over
{  SO(20+m-r; \IR)\times SO(4+m; \IR)}}~,
}
where $r$ is the reduction in rank relative to the toroidal compactification.

\subsec{$\IZ_2$}

This is the same automorphism used to construct the Type II-heterotic
dual pair in \schsenb. An example of a $K3$ surface which
admits this involution is the quartic polynomial in $\IC \IP^3$,
\eqn\quartic{
\sum_{i=1}^4 (z^i)^4 = 0~.
}
The $\IZ_2$ involution then acts as $z^1 \to -z^1$, $z^2 \to -z^2$,
$z^3 \to z^3$ and $z^4\to z^4$. This has eight fixed points on $K3$.
The involution interchanges eight pairs of anti-self-dual $(1,1)$-forms,
leaving the self-dual two-forms and the other anti-self-dual $(1,1)$-forms
invariant. To ensure this symmetry is freely acting on
$K3 \times S^1$ we accompany it by a $\pi$ rotation in the $S^1$.
Compactifying $d=11$ supergravity on $(K3\times S^1)/ \IZ_2$ yields
a theory with $N=2$ supersymmetry in six dimensions, with gauge
group $U(1)^{16}$ at a generic point.

To describe the action of the symmetry on the heterotic side,
first consider a point with enhanced $E_8\times E_8$ gauge symmetry.
It should be understood here and in the following subsections, that
when we work at a point of enhanced symmetry on the heterotic side
the dual will be a Type II theory compactified on some degenerate
$K3$ surface \refs{\witt,\aspinpt,\witstr}.
The $\IZ_2$ acts by interchanging the
two $E_8$ components of the Narain lattice, together with a half period
shift $\delta$ in the $\Gamma^{(1,1)}$ component of the lattice
corresponding to the $S^1$ of $d=11$ supergravity. The shift
must satisfy $\delta^2 =0$ for level-matching.
Generic points in the moduli space are reached by $SO(20,4)$
rotations that are compatible with the $\IZ_2$ action.
This is precisely the heterotic asymmetric orbifold construction
presented in \chpol. The massless spectrum
is the same as that found on the $d=11$ supergravity side at a
generic point in the moduli space.

\subsec{$\IZ_3$}

This symmetry has six fixed points on the $K3$. It acts on
$H^2(K3)$ by cyclic interchange of three groups of six
anti-self-dual $(1,1)$-forms, leaving the other two-forms invariant.
We will accompany this action on K3 by a $2\pi/3$ rotation in $S^1$.
The compactification of $d=11$ supergravity on $(K3\times S^1)/ \IZ_3$
has $N=2$ supersymmetry in six dimensions and gauge group $U(1)^{12}$
at generic points.

To describe the automorphism on the heterotic side
it is convenient to start at a point of enhanced gauge symmetry.
Consider, for example, the heterotic string compactified on a
Narain lattice $\Gamma^{(20,4)}$ with
$(D_1^6)_L^3 $$\times$$ (D_1)^2_L $$\times$$ (D_1)^4_R$
symmetry. Such a lattice is obtained from the Niemeier lattice $D_6^4$
with conjugacy classes given by even permutations of $(0,s,v,c)$, by
introducing a shift vector which decomposes one of the $D_6$ lattices
into $D_4$$\times$$D_1$$\times$$D_1$, followed by a flip of the signature
of the $D_4$ lattice, and shifts to give $(D_1)^4$ \lsw . In the
discussion here and below, $(s)$, $(v)$, and $(c)$, denote the
spinor, vector, and conjugate spinor,
conjugacy classes of a $D_n$ lattice. The $\IZ_3$
symmetry will act as a cyclic interchange of the three $(D_1)^6$ components
of the lattice, together with a shift $\delta$ by one-third of a period
in the $(D_1)_L\times (D_1)_R$ component corresponding to the
$S^1$ of the $d=11$ supergravity theory.
This orbifold action will satisfy level-matching provided $\delta^2=0$.
A generic point in the moduli space is then
reached by a $SO(20,4)$ rotation compatible with the $\IZ_3$ symmetry.
The massless spectrum of the resulting orbifold
agrees with that found on the supergravity side at generic points.

\subsec{$\IZ_4$}

Now we have four fixed points on $K3$ of order four, and four fixed
points of order two. Acting on $H^2(K3)$,
the symmetry cyclically interchanges four groups of four
two-forms, and changes the sign of two additional anti-self-dual
$(1,1)$-forms. The self-dual two-forms are invariant.
We accompany the action on K3
by a $\pi/2$ rotation in the $S^1$, to ensure the symmetry
is freely acting on $K3\times S^1$. Compactifying $d=11$
supergravity on this manifold will yield at generic points
a six-dimensional theory with $N=2$ supersymmetry and
$U(1)^{10}$ gauge symmetry.

To describe the action on the heterotic side, we begin with
an appropriate $\Gamma^{(20,4)}$ Narain lattice with
$((D_1)^4)^4 \times (D_1)^2_L \times
(D_1)^3_L$$\times$$(D_1)^3_R$ symmetry. A
symmetric combination of four $D_1$ lattices,
each of which is embedded in a different $(D_1)^4$ factor,
is purely right-moving.  The $\IZ_4$ symmetry then acts as cyclic
interchange of the four $(D_1)^4$
factors, as an involution, i.e., -1 on the first two $(D_1)_L$
factors, and as a shift $\delta$ by one-quarter of a period in a
$(D_1)_L\times (D_1)_R$ factor corresponding to the $S^1$
of $d=11$ supergravity. These isometries are easily identified
as subgroups of the isometry group of an embedding
$D_{16}$$\times$$D_2$ lattice, which can be obtained from the
$D_{24}$ Niemeier lattice, with conjugacy classes $(s)~ (v)$,
by flipping the signature of a $D_4$
factor together with necessary shifts. Level-matching is
satisfied for $\delta^2=0$.

Note that the symmetric embedding of a right-moving $D_1$ lattice
in four $(D_1)^4$ lattices ensures that it is left invariant under
the $\IZ_4$ interchange. A generic point in the moduli space is
reached by rotating the lattice by an element
of $SO(20,4)$ consistent with the $\IZ_4$ symmetry. The massless
spectrum of the resulting orbifold then agrees with that found
on the supergravity side.

\subsec{$\IZ_5$}

This automorphism has four fixed points on $K3$. Acting on $H^2(K3)$,
it cyclically permutes five groups of four two-forms.
The self-dual two-forms are invariant.
The action on K3 is combined with a $2\pi/5$ rotation
in the $S^1$. Compactifying $d=11$
supergravity on $(K3\times S^1)/\IZ_5$ yields a six-dimensional theory
with $N=2$ supersymmetry and $U(1)^{8}$ gauge symmetry at a generic
point.

On the heterotic side, we describe the action of the symmetry by
starting with a Narain lattice $\Gamma^{(20,4)}$
with $(D_1^4)^5_L \times (D_1)^4_R$ symmetry. Such a lattice is obtained
from the Niemeier lattice $D_4^6$ with conjugacy classes $(s,s,s,s,s,s)$,
$(0,[0,v,c,c,v])$ by flipping the signature of the first $D_4$ factor
and shifting to give $(D_1)^4_R$.
Here $[\cdots]$ denotes a sum over cyclic permutations.
The symmetry then simply acts as a cyclic interchange of the
five left-moving $(D_1)^4$ factors.  One may go to a complex
basis where the symmetry acts as a phase rotation. Two of the left-moving
complex bosonic coordinates are invariant under this symmetry. The left-moving
component of the compact direction corresponding to the $S^1$ of
$d=11$ supergravity is embedded in these two invariant complex coordinates.
The right-moving component corresponds to one of the invariant $(D_1)_R$
factors.
We accompany the action of the symmetry by a shift $\delta$ by $1/5$
of a lattice vector in this direction.
Level-matching is satisfied when $\delta^2=0$.
A generic point in the moduli space may be reached
by acting with a $SO(20,4)$ rotation compatible with the $\IZ_5$ symmetry.
The massless spectrum of the orbifold then coincides
with the supergravity side.

\subsec{$\IZ_6$}

This automorphism has a $\IZ_2$ and a $\IZ_3$ subgroup. The $\IZ_2$ subgroup
has six fixed points on the $K3$, the $\IZ_3$ subgroup has four fixed
points and the $\IZ_6$ has two fixed points. The action on
$H^2(K3)$ is as follows. Six groups of two (1,1)-forms
are cyclically interchanged by the $\IZ_6$, three groups of two
$(1,1)$-forms are cyclically interchanged by the $\IZ_3$,
and the two anti-self-dual $(1,1)$-forms pick up minus signs under the
$\IZ_2$ subgroup. The self-dual two-forms are invariant.
This group action is combined with a $\pi/3$ rotation
of the $S^1$. The compactification of $d=11$ supergravity on this
manifold yields a six-dimensional theory
with $N=2$ supersymmetry and $U(1)^{8}$ gauge symmetry at a generic
point.

On the heterotic side, we describe the action of the symmetry by
starting with an appropriate Narain lattice $\Gamma^{(20,4)}$
with $(D_1^2)^6_L \times (D_1^2)^3_L \times
(D_1)^2_L \times (D_1)^4_R$ symmetry.
The symmetry acts as a cyclic interchange of the first
six $(D_1)^2$ factors, cyclic interchange of the next three $(D_1)^2$
factors, and $-1$ on the remaining two left-moving $D_1$ factors,
together with a shift $\delta$ by $1/6$ of a lattice vector in the
direction corresponding to the $S^1$ of $d=11$ supergravity,
which is embedded in one of the invariant left-moving coordinates and
one of the $(D_1)_R$ factors.
The shift satisfies $\delta^2=0$. These isometries can be identified in
the isometry group of an embedding $D_{12}$$\times$$D_6$$\times$$D_2$
lattice, which embeds in $D_{24}$ as in the $\IZ_4$ example discussed
above. A generic point in the moduli space may be reached by acting with a
$SO(20,4)$ rotation preserving the $\IZ_6$ symmetry and the
massless spectrum of the orbifold then agrees with the supergravity side.

\subsec{$\IZ_7$}

Now we have three fixed points on $K3$. Acting on $H^2(K3)$,
the symmetry cyclically permutes seven groups of three two-forms,
and leaves one two-form invariant. A basis may be chosen
in which four two-forms are invariant under the action of this symmetry.
Three of these will correspond to linear combinations of the
self-dual $(2,0)$, $(0,2)$ and Kahler forms.
The action of this symmetry is to be combined with a $2\pi/7$ rotation
in the $S^1$. Compactifying $d=11$
supergravity on $(K3\times S^1)/\IZ_7$ yields at a generic point
a six-dimensional theory with $N=2$ supersymmetry and $U(1)^{6}$
gauge symmetry.

On the heterotic side, we describe the action of the symmetry by
starting with a Narain lattice $\Gamma^{(20,4)}$ with
$((D_1)^3)^7 \times (D_1)^3_R$ symmetry.
We obtain this lattice by beginning with the Niemeier lattice
$A_3^8$ with conjugacy classes $(3,[2,0,0,1,0,1,1])$, and flipping
the sign of an $A_3$$\times$$A_1$ sub-lattice followed by the
necessary shifts. The necessary sub-lattice is given by
the first $A_3$ factor together with an $A_1$ embedded
symmetrically in the remaining seven $A_3$ factors.
(Note that $SU(4)$ is isomorphic to $SO(6)$ so we can also
think of this as a $D_3^8$ lattice with a $Z_7$ isometry.)
The lattice is chosen such that a symmetric combination of seven
$D_1$ factors, with each $D_1$ lying in a different $(D_1)^3$ factor,
is purely right-moving. The symmetry acts as cyclic
interchange of the seven $(D_1)^3$ factors, together with a
shift $\delta$ by $1/7$ of a lattice vector in the direction
corresponding to the $S^1$ of the $d=11$ supergravity.
The translation corresponding to the $S^1$ is identified by pairing
a left-moving shift that acts symmetrically in all seven $(D_1)^3$
factors with a right-moving
shift in the $(D_1)^3_R$ lattice. Level-matching is satisfied when
$\delta^2=0$.  A generic point in the moduli space may be reached by
acting with a $SO(20,4)$ rotation compatible with the $\IZ_7$ symmetry
and the massless spectrum of the orbifold is then in agreement with
the supergravity side.

\subsec{$\IZ_8$}

This automorphism has a $\IZ_2$ and a $\IZ_4$ subgroup. The $\IZ_2$ subgroup
has four fixed points on the $K3$, the $\IZ_4$ subgroup has two fixed
points and the $\IZ_8$ has two fixed points. The action on $H^2(K3)$
is as follows. Eight groups of two two-forms
are cyclically interchanged by the $\IZ_8$, four groups of
two-forms are cyclically interchanged by the $\IZ_4$,
and one anti-self-dual $(1,1)$-form picks up a minus sign under the
$\IZ_2$ subgroup.  A basis
may therefore be chosen in which four two-forms are invariant.
Three of these will correspond to linear combinations of the
self-dual $(0,2)$, $(2,0)$ and Kahler forms.
This group action is combined with a $\pi/3$ rotation
of the $S^1$. The compactification of $d=11$ supergravity on
$(K3\times S^1)/\IZ_8$ gives a six-dimensional theory
with $N=2$ supersymmetry and $U(1)^{6}$ gauge symmetry at a generic
point in the moduli space.

On the heterotic side, we describe the action of the symmetry by
starting with a certain Narain lattice with signature $(20,4)$ and
$((D_1)^2)^8 \times (D_1)^4 \times (D_1)_L \times
(D_1)^3_R$ symmetry. A symmetric
combination of eight $D_1$ factors,
with each $D_1$ coming from the first component of each $(D_1)^2$ factor, and
four other $D_1$'s, each coming from a factor in the $(D_1)^4$ term,
is purely right-moving.
The symmetry then acts as a cyclic interchange of the
eight $(D_1)^2$ factors, cyclic interchange of the four $D_1$ factors
coming from the $(D_1)^4$ term,
and $-1$ on the last left-moving $D_1$ factor. This isometry
group is easily identified as a sub-group of the isometries of
$D_{16}$$\times$$D_4$, which can be obtained from $D_{24}$ as before.
The left-moving component of the direction corresponding to the $S^1$
of the $d=11$ supergravity
arises from the  symmetric combination of eight $D_1$ factors,
with each $D_1$ coming from the second component of each $(D_1)^2$ factor, and
four other $D_1$'s, each coming from a factor in the $(D_1)^4$ term.
The generator of the $\IZ_8$ is accompanied with a shift $\delta$ by
$1/8$ of a lattice vector in this direction.
The shift satisfies $\delta^2=0$.  A generic point in the moduli space
may be reached
by acting with a $SO(20,4)$ rotation preserving the $\IZ_8$ symmetry
and the massless spectrum of the orbifold agrees with the supergravity side.

\subsec{$(\IZ_2)^k$, $k=2,3,4$}

In this case we have $2^k-1$ $\IZ_2$ subgroups, each of which have eight
fixed points as in the $\IZ_2$ example above. Therefore this symmetry
group acts with a total of $8(2^k-1)$ fixed points. The action of the
symmetry may be represented by $k$ $\IZ_2$ generators, which we
denote $g_i$, $i=1,\cdots, k$. These
act on $H^2(K3)$ as follows. The self-dual two-forms
and three anti-self-dual two-forms are invariant. On the remaining
sixteen anti-self-dual $(1,1)$-forms, $g_1$ acts as
\eqn\gone{
( (-1)^8,1^8)~,
}
$g_2$ acts as
\eqn\gtwo{
( (-1)^4, 1^4, (-1)^4 , 1^4)~,
}
$g_3$ acts as
\eqn\gthree{
( (-1)^2, 1^2, (-1)^2 , 1^2, (-1)^2, 1^2, (-1)^2 , 1^2) ~,
}
and $g_4$ acts as
\eqn\gfour{
(-1,1,-1,1,-1,1,-1,1,-1,1,-1,1,-1,1,-1,1)~.
}
Here the superscripts denote repeated entries.

For $k \geq 2$, a shift in the additional $S^1$ of the $d=11$
supergravity theory is no longer sufficient to ensure the
symmetry is freely acting on $K3 \times S^1$. However, by compactifying
down to four dimensions on an additional torus with dimension $k-1$, i.e. on
$K3 \times T^{k}$, freely acting symmetries may be constructed as follows.
The momentum lattice describing the $T^k$ may be described by $k$ integers
$(m_i)$, with inner product $\sum m_i^2$.
We accompany the action of $g_i$, $i=1,\cdots,4$ by a shift with $m_i= \half$,
and the other components zero. The symmetry will then
act freely on $K3 \times T^k$. Compactifying $d=11$ supergravity on this
manifold yields a theory with $N=2$ supersymmetry in five dimensions
and gauge group $U(1)^{14}$ for $k=2$, $N=4$ supersymmetry in four
dimensions and gauge group $U(1)^{14}$ for $k=3$ and
$N=8$ supersymmetry in three dimensions and gauge group
$U(1)^{15}$ for $k=4$.

To describe the action of this symmetry on the heterotic side,
we start with an even self-dual $\Gamma^{19+k,3+k}$ Narain lattice
with $(D_1)_L^{16}\times  (D_1)_L^{3+k} \times (D_1)_R^{3+k}$ symmetry.
We can identify each of the $D_1$ factors in the $(D_1)_L^{16}$ with
one of the sixteen $(1,1)$-forms of the $K3$ on which the symmetry
acts in a nontrivial way as described above. The action of the
$g_i$ on these factors then may be identified with the action on the
$(1,1)$-forms. Likewise we identify the accompanying shifts on the $T^k$ of
the $d=11$ supergravity theory with the same left-right symmetric shifts on the
$(D_1)_L \times (D_1)_R$ factors. The massless spectrum of the resulting
orbifolds then coincides with that on the supergravity side.

At this point we note that the $(\IZ_2)^2$ and $(\IZ_2)^3$ asymmetric
orbifolds described here recover all of the rank $-$$12$ and
rank $-$$14$ solutions found in the fermionic construction \refs{\chl,\chpol}.
The clearest evidence for this is the existence of $\IZ_2$,
$(\IZ_2)^2$, and $(\IZ_2)^3$ orbifolds containing enhanced symmetry
points with, respectively, Kac-Moody level $2$, level $4$, and level
$8$ realizations of the gauge symmetry. Each $\IZ_2$ cyclic automorphism
of repeated factor groups doubles the Kac-Moody level.
Thus, for example, beginning with the Narain lattice
$E_8$$\times$$E_8$$\times$$(D_2)^3$ we obtain reduced rank models
with, respectively,
$(E_8)_2$$\times$$(SO(4))^3$, $SO(8)_4$$\times$$(SO(4))^3$,
and $SO(4)_8$$\times$$(SO(4))^3$ gauge symmetry. Here the
subscripts denote the Kac-Moody level. These
solutions have already been obtained in the fermionic construction.

\subsec{$(\IZ_3)^2$}

In this case, we have four $\IZ_3$ subgroups, each of which is
the stationary subgroup for six fixed points. The symmetry
is generated by two $\IZ_3$ generators which we label $g_1$ and $g_2$.
The symmetry leaves the self-dual two-forms invariant.
The generators act on eighteen of the anti-self-dual $(1,1)$-forms as follows.
$g_1$ cyclically interchanges three groups of six $(1,1)$-forms.
Within each of these three groups, $g_2$ acts by cyclically interchanging
three sets of two $(1,1)$-forms. In order that the symmetry be
freely acting on the $d=11$ supergravity side, we consider
compactification on $(K3 \times S^1\times S^1)/(\IZ_3)^2$, where each generator
$g_1$, $g_2$ is accompanied by a $2\pi/3$ rotation in the first and
second circle respectively. This yields a theory with $N=2$ supersymmetry
in five dimensions and gauge group $U(1)^{10}$.

On the heterotic side we start by describing the action of this symmetry
on a Narain lattice $\Gamma^{(21,5)}$
with symmetry $ ((D_1^2)^3)^3_L \times (D_1)^3_L
\times (D_1)^5_R$. The action on the $((D_1^2)^3)^3_L$ components of the
lattice are completely analogous to the action of the symmetry on the
cohomology. Likewise the generators of the symmetry are accompanied
by the shifts described above, acting on the corresponding
$(D_1)_L \times (D_1)_R$ factors. Performing a $SO(21,5)$ rotation
of the lattice compatible with this symmetry yields an orbifold theory
with the gauge group broken down to $U(1)^{10}$ and $N=2$ supersymmetry
in five dimensions, as on the supergravity side.

\subsec{$\IZ_2 \times \IZ_4$}

This case is very similar to the $\IZ_4$ and $\IZ_2$ cases.
The symmetry acts with two groups of eight
fixed points of order two, and two groups of four fixed points of order four.
Let $g_1$ denote the generator of the $\IZ_2$ and $g_2$ denote the
generator of the $\IZ_4$.
The action of $g_2$ on $H^2(K3)$ is the same as in the
$\IZ_4$ case. Recall, this involved cyclically interchanging
four groups of four two-forms which we will
denote by $\omega_{ijk}$ with $i=1,2$, $j=1,2$ and
$k=1,\cdots,4$. $g_2$ also acts by changing the
sign of two anti-self-dual $(1,1)$-forms $\omega_1$ and $\omega_2$.
$g_1$ acts by changing the sign of two of $\omega_1$, $\omega_2$ and
two other anti-self-dual $(1,1)$-forms $\omega_3$ and $\omega_4$.
In addition, $g_1$ interchanges $\omega_{2jk}$ with $k=1,2$ with
$\omega_{2jk'}$ with $k'=3,4$ respectively.
In order that the symmetry be freely acting on the $d=11$ supergravity side,
we consider
compactification on $(K3 \times S^1\times S^1)/(\IZ_2 \times \IZ_4)$,
where $g_1$ is accompanied by a $\pi$ rotation in one circle, and
$g_2$ is accompanied by a $\pi/2$ rotation in the other.
This yields a theory with $N=2$ supersymmetry
in five dimensions and gauge group $U(1)^{10}$.
The action of the symmetry on the heterotic string compactified
on a $\Gamma^{(21,5)}$ lattice follows in an obvious way given the
action on the cohomology and the previous description of the
$\IZ_2$ and $\IZ_4$ cases.

\subsec{$(\IZ_4)^2$}

The symmetry acts on the $K3$ surface with six groups of four fixed points
of order four.
Let us denote the generator of each $\IZ_4$ by $g_1$ and $g_2$ respectively.
To describe the action of this symmetry it is convenient to introduce
two two-forms in addition to the 22 two-forms arising from the $K3$.
These additional two-forms may be thought of as arising from
$H^2(K3\times T^2)$. The reason we introduce these extra two-forms
is that the symmetry may then be written as a combination of cyclic
interchanges of linear combinations of the 24 two-forms, which
we will denote by $\omega_{ij}$, with $i=1,\cdots,6$ and $j=1,\cdots,4$.
The linear combinations are of course chosen in such a way that the
additional two-forms are invariant under the symmetry.
$g_1$ acts by cyclically interchanging on the $j$ index, the
$\omega_{ij}$ with $i=1,\cdots,4$ and interchanges $\omega_{6j}$ with
$j=1,2$ with $\omega_{6j'}$ with $j'=3,4$.
$g_2$ acts by cyclically interchanging on the $j$ index, the
$\omega_{ij}$ with $i=3,\cdots,6$ (with the interchange of $\omega_{4j}$ being
in the reverse order relative to the $g_1$ action)
and interchanges $\omega_{2j}$ with $j=1,2$ with $\omega_{2j'}$ with $j'=3,4$.
$g_1$ and $g_2$ leave invariant the
self-dual two-forms and one anti-self-dual two-form of $K3$.
If we accompany $g_1$ with a $\pi/2$ rotation of one circle, and
$g_2$ with a $\pi/2$ rotation of another circle, then
the symmetry acts freely on $K3\times S^1\times S^1$.
Compactifying $d=11$ supergravity on $(K3\times S^1\times S^1)/(\IZ_4)^2$
then yields a theory with $N=2$ supersymmetry in five dimensions
and gauge group $U(1)^{8}$. The action of the symmetry on the
heterotic side follows straightforwardly from the action on the
cohomology given above, and the single $\IZ_4$ case worked out above.

\subsec{$\IZ_2 \times \IZ_6$}

Let us denote the generator of $\IZ_2$ by $g_1$ and the generator
of $\IZ_6$ by $g_2$.
The symmetry acts on $K3$ with three groups of six fixed points
of order two, and three groups of two fixed points of order six.
To describe the action of the symmetry in a convenient way,
consider the basis of two-forms $\omega_{ij}$, with $i=1,2,3$ and
$j=1,\cdots,6$, $\omega_k$ with $k=1,2,3$ and one self-dual
two-form $\omega'$. $g_1$ acts by changing the sign of
$\omega_2$ and $\omega_3$ and interchanges $\omega_{2j}$ with
$j=1,2,3$ with $\omega_{2j'}$ for $j'=4,5,6$, respectively.
$g_1$ also acts by interchanging $\omega_{3j}$ with
$j=1,2,3$ with $\omega_{3j'}$ for $j'=4,5,6$, respectively.
On $H^2(K3)$, $g_2$ changes the sign of $\omega_1$ and $\omega_2$,
cyclically interchanges $\omega_{1j}$ and $\omega_{2j}$ on the $j$ index,
and acts with a $\IZ_3$ cyclic permutation of the pairs
$\omega_{3j}$ ($j=1,2$), $\omega_{3j'}$ ($j'=3,4$) and
$\omega_{3j''}$ ($j''=5,6$). Both $g_1$ and $g_2$ leave the
self-dual two-forms invariant, and one linear combination
of the anti-self-dual two-forms.
Accompanying $g_1$ with a $\pi$ rotation of one circle and
$g_2$ with a $\pi/3$ rotation of another circle, the symmetry
is freely on $K3\times S^1\times S^1$.
The compactification of $d=11$ supergravity on $(K3\times S^1\times S^1)/
(\IZ_2\times \IZ_6)$ gives a theory in five dimensions with $N=2$
and gauge group $U(1)^{8}$. The action of the symmetry on the
heterotic side follows straightforwardly from the previous examples.

\newsec{Conclusions}

The theories we have constructed should have interesting
implications for S-duality in four dimensions. As has been
pointed out \refs{\duff,\witt} string-string duality in
six dimensions between the Type IIA theory compactified on $K3$ and
the heterotic string on a four-torus implies S-duality for the theories further
compactified to four dimensions on a torus. This follows
since string-string duality in six dimensions maps T-duality of one
theory to S-duality of the dual. Provided T-duality is not modified
quantum mechanically, S-duality of the dual theory follows.
A similar story should be true for the orbifold theories
described in this paper, although now, in general, only a subgroup
of $SL(2,\IZ)_T$ and likewise $SL(2,\IZ)_S$ will commute with the
orbifolding procedure.
This has already been observed in the context of a five-dimensional
maximally supersymmetric dual pair, further compactified to
four dimensions in \vafawit.  Related phenomena in the context
of Type II-Type II dual pairs have been considered in \senvafa.

As has been pointed out \chpol, the elements of the
original $SL(2,\IZ)_S$ of the toroidally compactified
theory which do not commute with the orbifolding, are
expected not to leave individual points in the moduli space
invariant. In particular, in the $\IZ_2$ orbifold theory
considered in \chpol, such elements are expected to exchange
points in the moduli space corresponding to dual
electric-magnetic pairs. The appearance of dual gauge
groups in the same moduli space provides further
evidence for S-duality. This phenomenon is expected to generalize in the
theories described in this paper, where the non-commuting
duality elements act non-trivially within the moduli space.
It would be interesting to classify the enhanced
symmetry points in our new heterotic compactifications and to
study these duality transformations more carefully.

Finally, we note that the maximally supersymmetric models we have constructed
provide a new starting point for obtaining dual pairs with reduced
supersymmetry. Type II-heterotic dual pairs with partially broken
supersymmetry have recently been studied in \refs{\fhsv, \vafawit, \hls}.

\bigskip
{\bf Acknowledgements}

We thank P. Aspinwall, J. Harvey, J. Lykken, J. Polchinski and A. Strominger
for useful conversations.  This work was supported in part by
NSF Grants PHY 91-16964 and PHY 94-07194.

\listrefs
\end